# Possible Alternate Scenario for short Duration GRBs

C Sivaram and Kenath Arun

Indian Institute of Astrophysics, Bangalore

**Abstract:** In this paper we look at new class of objects made up entirely of dark matter particles. We look at these objects as possible candidate for short duration gamma ray bursts eliminating the baryon load problem. These could also provide a possible scenario for the formation of sub-stellar black holes, distinct from the usual Hawking black hole.

Short gamma ray bursts are those gamma ray bursts that have a shorter duration (<0.2 – 2 s) and a harder spectrum as compared to the duration of 2 – 200 s for long GRBs. Short GRBs are possibly due to the merger of two neutron stars, where as, the long GRBs are due to the collapse of very massive stars. The spectrum observed is harder because the objects merging to produce the GRB are more compact. In the case of short duration GRB, the energy released is the binding energy of the neutron stars which is of the order of $\sim 10^{53}\, ergs$.

Most sources capable of impulsively releasing the $10^{53}\, ergs$ or more of energy required to power a GRB, however, contain so much matter around them that if the energy released were used to accelerate even a very small fraction $\left(\sim 10^{-3}\right)$ of the baryons present, only a non-relativistic wind would result. This is known as the baryon-loading problem. [1, 2]

It has been hoped that the geometry of the sources is such that at least some of the energy released is channelled along directions relatively free of baryons, so that relativistic bulk motion and the ensuing beaming of radiation may occur along certain lines of sight. So far, this has not yet been fully demonstrated for any theoretical source of GRBs. [3, 4, 5]

Here we discuss a new class of objects made of pure dark matter particles. If these dark matter particles (of mass $m_D \sim 10 GeV$ to $1 TeV$) cluster and form gravitationally bound objects, these pairs of dark matter particles can annihilate throughout these objects. These dark matter particle-antiparticle pairs can undergo annihilation and produce high energy gamma rays which could be detected. These high energy gamma rays could be a signature of this new class of objects. [6]

The Chandrasekhar mass (upper limit) for these degenerate DM objects is given by:

$$M_{D(CH)} = \left(\frac{\hbar c}{G}\right)^{3/2} \frac{1}{m_D^2} \qquad \ldots (1)$$

For a dark matter particle of mass $m_D \sim 100 GeV$, this works out to be:

$$M_D \approx 10^{27} g = 10^{-6} M_{sun} \qquad \ldots (2)$$

The size of these objects is given by (for the usual degenerate gas configuration; thermal energy not being relevant):

$$M_D^{1/3} R = \frac{92 \hbar^2}{G m_D^{8/3}} \qquad \ldots (3)$$

For the $10^{-6}$ solar mass object the size works out to be:

$$R \approx 10^5 cm \qquad \ldots (4)$$

If their mass exceeds this limit, they will collapse to form black holes of size given by:

$$R_S = \frac{2GM}{c^2} \approx 1 cm \qquad \ldots (5)$$

(For DM particles of mass $m_D = 100 GeV$)

The energy released during the collapse is given by:

$$E = \frac{GM^2}{R} \approx 10^{48} ergs \qquad \ldots (6)$$

This energy is released in the form of gravitational waves.

If equal amount of baryonic matter collapses along with the dark matter to form the black hole, then the baryonic matter will be heated up to a temperature *T*, according to:

$$M R_g T = 10^{48} ergs \qquad \ldots (7)$$

This gives $T \approx 10^{12} K$, $R_g$ being the universal gas constant.

This energy corresponds to gamma ray frequencies. Since the mass heated up is $\sim 10^{-6}$ solar mass, in this scenario, the 'Baryon Load' problem seems ameliorated as the relativistic kinetic energy corresponds to Lorentz factor of $\sim 10^2 - 10^3$.

The time scale of the gamma ray burst here is given by:

$$t_{burst} = \sqrt{\frac{R^3}{GM}} \approx 0.01s \qquad \ldots (8)$$

The matter will expand to $> 10^9 m$ in a few seconds. Depending on the ambient medium there could be afterglows. The expansion would cause lowering of the temperature, resulting in production of X-rays, UV, etc, that is, radiation of successively longer wavelengths over longer intervals of time.

The peak wavelength would scale with the expansion time scale as roughly $\lambda \sim t^{-1}$, so that a few days after the initial burst, the wavelength would be in the ultraviolet to visible but with intensity far less (by a factor of $10^4$) than the initial luminosity.

As in this scenario, magnetic fields are not expected to be present [6], the radiation would not be polarised like in some GRB sources. This could be an alternative scenario for short duration $(0.1 - 0.01s)$ sub-luminous gamma ray bursts.

Again in this scenario, unlike in some other models of short duration GRB's we do expect much lower fluxes of neutrinos and gravitational waves to be simultaneously emitted (for details see ref. [6]). This could be another distinct signature of this model.

As an additional consequence this could be another way in which primordial black holes (< stellar mass) can form, apart from Hawking black holes. The masses of these sub-stellar mass black holes depend on the mass of the dark matter particles given by equation (1). For different dark matter particle masses the black hole mass is given in the following table: (see also ref. [6])

| $m_D(GeV)$ | $M_D(g)$ |
|---|---|
| 10 | $10^{29}$ |
| 100 | $10^{27}$ |
| 250 | $4 \times 10^{25}$ |
| 500 | $10^{25}$ |
| 1000 | $3 \times 10^{24}$ |

These black holes will then evaporate due to the usual Hawking radiation with a life time given by: $\tau_{ev} = \dfrac{5120\pi G^2 M^3}{\hbar c^4}$ … (9)

For this lifetime to be of the order of the Hubble time, the mass of the black hole should be: $M \approx 10^{14} g$ … (10)

This implies that we would expect all of the above black hole masses (formed by the collapse of DM dominated objects) to be still present at the present epoch of the universe.

**Concluding Remarks:** This article presents an alternate scenario for short duration gamma ray bursts due to collapse of dark matter dominated objects. This scenario successfully eliminates the 'Baryon Load Problem'. The remnant of these GRBs will be black holes of sub-stellar mass, which provides another mechanism for the formation of such (primordial) black holes, apart from the usual Hawking black holes.

**Reference:**

1. Kluzniak W, Ap. J., 508, L29, 1998
2. Piran T. 1997, in Unsolved Problems in Astrophysics, ed. J. N. Bahcall & J. P. Ostriker (Princeton: Princeton Univ. Press), 343
3. Rasio F., & Shapiro S. L. 1992, Ap. J., 401, 226
4. Kluzniak W., & Lee W. H. 1998, Ap. J., 494, L53
5. Wilson J. R. *et al*. 1998, in AIP Conf. Proc. 4, Gamma-Ray Bursts, Ed. C. A. Meegan *et al*. (New York: AIP), 788
6. Sivaram C and Kenath Arun, arXiv:0910.2306v1 [astro-ph.CO], October 2009